\documentclass{article}
\usepackage{amssymb}
\begin{document}
\title{Mapping of solutions of the Hamilton--Jacobi equation by an arbitrary canonical transformation}

\author{G.F.\ Torres del Castillo \\ Departamento de F\'isica Matem\'atica, Instituto de Ciencias \\
Universidad Aut\'onoma de Puebla, 72570 Puebla, Pue., M\'exico \\[2ex]
H.H.\ Cruz Dom\'inguez, A.\ de Yta Hern\'andez, J.E.\ Herrera Flores, \\
and A.\ Sierra Mart\'inez \\ Facultad de Ciencias F\'isico Matem\'aticas \\ Universidad Aut\'onoma de Puebla, Apartado postal 165 \\ 72001 Puebla, Pue., M\'exico}

\maketitle

\begin{abstract}
It is shown that given an arbitrary canonical transformation and an arbitrary Hamiltonian, there is a naturally defined mapping that sends any solution of the Hamilton--Jacobi (HJ) equation into a solution of the HJ equation corresponding to the new Hamiltonian.
\end{abstract}


\section{Introduction}
In the framework of classical mechanics, the Hamilton--Jacobi (HJ) equation constitutes a very useful tool in the solution of the equations of motion. For a given Hamiltonian, any complete solution of the corresponding HJ equation yields the general solution of the Hamilton equations (see, e.g., Ref.\ 1). The form of the Hamilton equations is preserved by any canonical transformation, but the expression of the Hamiltonian may be modified and, therefore, the form of the HJ equation is also affected by a canonical transformation. A natural question is: how can we relate the solutions of the HJ equations corresponding to the two Hamiltonians?

As remarked in Ref.\ 2, in a canonical transformation, the original coordinates, $q_{i}$, may be functions of the new coordinates, the new momenta, and the time, $q_{i} = q_{i}(Q_{j}, P_{j}, t)$, whereas a solution, $S(q_{i}, t)$, of the HJ equation depends on the coordinates and the time only, so that, a solution of the HJ equation corresponding to the new Hamiltonian cannot be obtained from $S(q_{i}, t)$ by simply substituting the $q_{i}$ as functions of $Q_{i}, P_{i}$, and $t$.

In Ref.\ 2 the required relation was found in the restricted case where the canonical transformation does not involve the time and the Hamiltonians do not depend on the time. Furthermore, the solutions of the HJ equations considered there are of the form $S(q_{i}, t) = W(q_{i}) - Et$, which do exist for time-independent Hamiltonians, but form a restricted class of solutions.

In Ref.\ 3 it was shown that given a Hamiltonian that can depend on the time and a {\em one-parameter group}\/ of canonical transformations that may involve the time, the action of the group of transformations on a solution of the HJ equation can be defined by means of a partial differential equation analogous to the HJ equation, with the generating function of the transformations in place of the Hamiltonian, but the effect of a single canonical transformation was not determined there.

In this paper we find the effect of an arbitrary canonical transformation (not necessarily an element of a continuous group of transformations) on an arbitrary solution of the HJ equation (not necessarily complete) corresponding to a Hamiltonian that can depend on the time. As we shall show below, this effect can be derived in a simple manner and the results of Refs.\ 2 and 3 are readily reproduced from the general expression obtained here.

In Section 2 the main results are established and in Section 3 several examples are given.

\section{Solutions of the HJ equation and canonical transformations}
The HJ equation corresponding to a given Hamiltonian of a system with $n$ degrees of freedom, $H(q_{i}, p_{i}, t)$, is the partial differential equation
\begin{equation}
H \left( q_{i}, \frac{\partial S}{\partial q_{i}}, t \right) + \frac{\partial S}{\partial t} = 0. \label{hje}
\end{equation}
If we perform a canonical transformation, $Q_{i} = Q_{i}(q_{j}, p_{j}, t)$, $P_{i} = P_{i}(q_{j}, p_{j}, t)$, the Hamiltonian $H(q_{i}, p_{i}, t)$ has to be replaced by a new one, $K(Q_{i}, P_{i}, t)$, which gives rise to another HJ equation
\begin{equation}
K \left( Q_{i}, \frac{\partial S'}{\partial Q_{i}}, t \right) + \frac{\partial S'}{\partial t} = 0, \label{nhje}
\end{equation}
and we want to find a way of constructing a solution $S'(Q_{i}, t)$ of Eq.\ (\ref{nhje}) from each solution $S(q_{i}, t)$ of Eq.\ (\ref{hje}). To this end, we start by pointing out a useful characterization of the solutions of the HJ equation (cf.\ also Ref.\ 4).

\noindent {\bf Proposition 1.} Any solution, $S(q_{i}, t)$, of the HJ equation (\ref{hje}) defines a surface (a submanifold) of the extended phase space, given by
\begin{equation}
p_{i} = \frac{\partial S}{\partial q_{i}}, \label{subvar}
\end{equation}
$i =  1, 2, \ldots, n$, on which the linear differential form $p_{i} {\rm d} q_{i} - H {\rm d} t$ is exact (with summation over repeated indices); in fact,
\begin{equation}
p_{i} {\rm d} q_{i} - H {\rm d} t = {\rm d} S. \label{ds}
\end{equation}
Conversely, a submanifold of the extended phase space, given by $n$ functions
\begin{equation}
p_{i} = F_{i}(q_{j}, t),
\end{equation}
on which the differential form $p_{i} {\rm d} q_{i} - H {\rm d} t$ is exact, defines (up to an additive constant) a solution of the HJ equation. (The solution is the function $S$ determined by Eq.\ (\ref{ds}).)

It may be noticed that the function $S$ appearing in Eqs.\ (\ref{hje}), (\ref{subvar}), and (\ref{ds}) may contain some parameters (as in the case of a complete solution), but this is not essential at this point. For example, if the Hamiltonian is taken as
\begin{equation}
H = \frac{p^{2}}{2m} - ktq, \label{varf}
\end{equation}
where $k$ is a constant, then, on the two-dimensional submanifold of the extended phase space defined by
\[
p = \frac{kt^{2}}{2},
\]
we have
\begin{eqnarray*}
p {\rm d} q - H {\rm d} t & = & \frac{kt^{2}}{2} {\rm d} q - \left( \frac{k^{2} t^{4}}{8m} - ktq \right) {\rm d} t \\
& = & {\rm d} \left( \frac{kt^{2}q}{2} - \frac{k^{2}t^{5}}{40 m} \right).
\end{eqnarray*}
Hence, the function
\begin{equation}
S = \frac{kt^{2}q}{2} - \frac{k^{2}t^{5}}{40 m} \label{solhj}
\end{equation}
is a solution, without arbitrary parameters, of the HJ equation for the Hamiltonian (\ref{varf}). However, for each value of the constant $a$, the equation
\begin{equation}
p = \frac{kt^{2}}{2} + a, \label{fol}
\end{equation}
defines a two-dimensional submanifold of the extended phase space, on which the differential form $p {\rm d} q - H {\rm d} t$ is exact. In this case one obtains a complete solution of the HJ equation, which is related to the fact that the collection of submanifolds (\ref{fol}) covers all the extended phase space.

On the other hand, the coordinate transformation
\begin{equation}
Q_{i} = Q_{i}(q_{j}, p_{j}, t), \qquad P_{i} = P_{i}(q_{j}, p_{j}, t), \label{coordt}
\end{equation}
is canonical if and only if there exists some function $F$ such that
\begin{equation}
P_{i} {\rm d} Q_{i} - K {\rm d} t - (p_{i} {\rm d} q_{i} - H {\rm d} t) = {\rm d} F, \label{ct}
\end{equation}
where $K$ is the new Hamiltonian.

For instance, the transformation
\begin{equation}
Q = q - Vt, \qquad P = p - mV, \label{gal}
\end{equation}
where $m$ and $V$ are constants, is canonical since
\begin{eqnarray*}
P {\rm d} Q - K {\rm d} t - (p {\rm d} q - H {\rm d} t) & = & (p - mV) ({\rm d} q - V {\rm d} t) - p {\rm d} q + (H - K) {\rm d} t \\
& = & - Vp {\rm d} t - mV {\rm d} q + mV^{2} {\rm d} t + (H - K) {\rm d} t \\
& = & {\rm d} (- mVq) + (- Vp + mV^{2} + H - K) {\rm d} t,
\end{eqnarray*}
which shows that the new Hamiltonian must be
\begin{equation}
K = H - Vp + mV^{2} + \phi(t) = H - VP + \phi(t), \label{k}
\end{equation}
where $\phi(t)$ is an arbitrary function of $t$ only, and
\begin{equation}
F = - mVq - \int^{t} \phi(u) {\rm d} u. \label{efe}
\end{equation}

Whereas the differential form $p_{i} {\rm d} q_{i} - H {\rm d} t$ (and, similarly, $P_{i} {\rm d} Q_{i} - K {\rm d} t$) is exact only on some submanifolds of the extended phase space (of dimension not greater than $n + 1$), the combination $P_{i} {\rm d} Q_{i} - K {\rm d} t - (p_{i} {\rm d} q_{i} - H {\rm d} t)$ is exact in open neighborhoods of the extended phase space (that is, in $(2n + 1)$-dimensional regions). Hence, from Eq.\ (\ref{ct}) we see that, if $p_{i} {\rm d} q_{i} - H {\rm d} t$ is an exact differential on some submanifold of the extended phase space, then $P_{i} {\rm d} Q_{i} - K {\rm d} t$ is also exact on that submanifold. Thus, if $S(q_{i}, t)$ is a solution of the HJ equation (\ref{hje}), then
\begin{equation}
P_{i} {\rm d} Q_{i} - K {\rm d} t = {\rm d} (S + F) \label{npcf}
\end{equation}
on the submanifold (\ref{subvar}) and, according to Proposition 1,
\begin{equation}
S' = S + F \label{sprime}
\end{equation}
is a solution of the HJ equation (\ref{nhje}), provided that the right-hand side of Eq.\ (\ref{sprime}) is expressed in terms of the $Q_{i}$ and $t$, eliminating the other variables by means of Eqs.\ (\ref{subvar}) and (\ref{coordt}).

For instance, in the case of the Hamiltonian (\ref{varf}) and the canonical transformation (\ref{gal}), from Eq.\ (\ref{k}) we find that, {\em choosing}\/ $\phi(t) = kVt^{2} - mV^{2}/2$,
\begin{equation}
K = \frac{P^{2}}{2m} - ktQ, \label{nvarf}
\end{equation}
which has the form of the original Hamiltonian (\ref{varf}), with only $q$ and $p$ replaced by $Q$ and $P$, respectively, and from (\ref{efe}),
\begin{equation}
F = - mVq - \frac{kVt^{3}}{3} + \frac{mV^{2}t}{2}. \label{fngen}
\end{equation}
Thus, making use of Eqs.\ (\ref{sprime}), (\ref{solhj}), and (\ref{gal}),
\begin{eqnarray}
S' & = & \frac{kt^{2}q}{2} - \frac{k^{2}t^{5}}{40 m} - mVq - \frac{kVt^{3}}{3} + \frac{mV^{2}t}{2} \nonumber \\
& = & \frac{kt^{2}Q}{2} - \frac{k^{2}t^{5}}{40 m} - mVQ + \frac{kVt^{3}}{6} - \frac{mV^{2}t}{2}, \label{exasprime}
\end{eqnarray}
which is a solution of the HJ equation corresponding to the Hamiltonian (\ref{nvarf}); $S'$ contains the parameter $V$ and is a complete solution.

\subsection{Connection with previous results}
In the case where the Hamiltonian $H$ does not depend on $t$ and the canonical transformation (\ref{coordt}) does not involve the time, choosing $K = H$, the function $F$, on the right-hand side of Eq.\ (\ref{ct}), does not depend on $t$, then, making use of the fact that the HJ equation (\ref{hje}) admits solutions of the form
\begin{equation}
S(q_{i}, t) = W(q_{i}) - Et, \label{sep}
\end{equation}
where $E$ is a constant, from Eq.\ (\ref{sprime}) we obtain a solution of the HJ equation (\ref{nhje}),
\[
S' = W + F - Et,
\]
which is also of the form (\ref{sep}), $S' = W' - Et$, with
\[
W' = W + F,
\]
as given in Eq.\ (14) of Ref.\ 2.

As we shall show, in the case where the Hamiltonian $H$ may depend on $t$ and we have a one-parameter group of canonical transformations generated by some function $G$ defined on the extended phase space, the function $S'$ given by Eq.\ (\ref{sprime}) satisfies the partial differential equation
\begin{equation}
G \left( Q_{i}, \frac{\partial S'}{\partial Q_{i}}, t \right) + \frac{\partial S'}{\partial \alpha} = 0, \label{hjl}
\end{equation}
where $\alpha$ is the parameter of the group, with the initial condition $S'|_{\alpha = 0} = S$ (assuming that for $\alpha = 0$ the canonical transformation generated by $G$ reduces to the identity). In Ref.\ 3, the action of a one-parameter group of canonical transformations on a solution of the HJ equation was {\em defined}\/ by Eq.\ (\ref{hjl}).

In order to derive Eq.\ (\ref{hjl}), we note that if both sides of Eq.\ (\ref{ct}) (including the original coordinates $q_{i}, p_{i}$) are expressed as functions of $Q_{i}, P_{i}, t$, and $\alpha$, then, taking the partial derivative with respect to $\alpha$,
\[
- \left( \frac{\partial K}{\partial \alpha} \right)_{Q,P,t} {\rm d} t - \left( \frac{\partial p_{i}}{\partial \alpha} \right)_{Q,P,t} {\rm d} q_{i} - p_{i} {\rm d} \left( \frac{\partial q_{i}}{\partial \alpha} \right)_{Q,P,t} + \left( \frac{\partial H}{\partial \alpha} \right)_{Q,P,t} {\rm d} t = {\rm d} \left( \frac{\partial F}{\partial \alpha} \right)_{Q,P,t},
\]
where we have made use of the notation $(\partial/\partial \alpha)_{Q,P,t}$ to emphasize that $Q_{i}, P_{i}$, and $t$ are held constant in the differentiation. Thus,
\[
\left( \frac{\partial (H - K)}{\partial \alpha} \right)_{Q,P,t} {\rm d} t - \left( \frac{\partial p_{i}}{\partial \alpha} \right)_{Q,P,t} {\rm d} q_{i} + \left( \frac{\partial q_{i}}{\partial \alpha} \right)_{Q,P,t} {\rm d} p_{i} = {\rm d} \left[ \left( \frac{\partial F}{\partial \alpha} \right)_{Q,P,t} + p_{i} \left( \frac{\partial q_{i}}{\partial \alpha} \right)_{Q,P,t} \right].
\]
Letting
\begin{equation}
G \equiv - \left( \frac{\partial F}{\partial \alpha} \right)_{Q,P,t} - p_{i} \left( \frac{\partial q_{i}}{\partial \alpha} \right)_{Q,P,t} \label{infgene}
\end{equation}
we have
\[
\left( \frac{\partial p_{i}}{\partial \alpha} \right)_{Q,P,t} = \frac{\partial G}{\partial q_{i}}, \qquad \left( \frac{\partial q_{i}}{\partial \alpha} \right)_{Q,P,t} = - \frac{\partial G}{\partial p_{i}},
\]
hence, making use of Eq.\ (\ref{sprime}), we have
\begin{eqnarray*}
\left( \frac{\partial S'}{\partial \alpha} \right)_{Q,P,t} & = & \frac{\partial S}{\partial q_{i}} \left( \frac{\partial q_{i}}{\partial \alpha} \right)_{Q,P,t} + \left( \frac{\partial F}{\partial \alpha} \right)_{Q,P,t} \\
& = & p_{i} \left( \frac{\partial q_{i}}{\partial \alpha} \right)_{Q,P,t} + \left( \frac{\partial F}{\partial \alpha} \right)_{Q,P,t} \\
& = & - G,
\end{eqnarray*}
as was to be shown. (Compare with the derivation given in Ref.\ 2, for the time-independent case.)

In the example considered above, the transformations (\ref{gal}) form a one-parameter group with the parameter being $V$. According to Eqs.\ (\ref{gal}), (\ref{fngen}) and (\ref{infgene}), the generating function of the group is $G(Q, P, t) = mQ - tP + kt^{3}/3$, and one readily verifies that the expression (\ref{exasprime}) satisfies Eq.\ (\ref{hjl}). Since $K$ and $H$ have the same form, replacing $Q$ by $q$ in the expression (\ref{exasprime}) one obtains a complete solution of the HJ equation for $H$.

\section{Further examples}
In this section we give two additional examples of the use of Eq.\ (\ref{sprime}). We begin with the Hamiltonian
\[
H = {\rm e}^{- 2 \gamma t} \frac{p^{2}}{2m} + {\rm e}^{2 \gamma t} \frac{m \omega^{2} q^{2}}{2},
\]
where $\gamma$ is a constant, which corresponds to a damped harmonic oscillator. The coordinate transformation
\[
Q = {\rm e}^{\gamma t} q, \qquad P = {\rm e}^{- \gamma t} p
\]
is canonical and from Eq.\ (\ref{ct}) one finds that the new Hamiltonian can be taken as
\[
K  = \frac{P^{2}}{2 m} + \frac{m \omega^{2} Q^{2}}{2} + \gamma PQ,
\]
with $F = 0$. By contrast with $H$, the Hamiltonian $K$ does not depend explicitly on $t$ and therefore the HJ equation for $K$ admits separable solutions of the form
\[
S' = - E t + f(Q),
\]
where $E$ is a separation constant and $f$ satisfies
\[
\frac{{\rm d} f}{{\rm d} Q} = - m\gamma Q \pm \sqrt{2 mE - m^{2} (\omega^{2} - \gamma^{2}) Q^{2}},
\]
thus, a (complete) solution of the HJ equation for $H$ is given by
\[
S = - Et - \frac{1}{2} m \gamma {\rm e}^{2 \gamma t} q^{2} \pm \int^{{\rm e}^{\gamma t} q}  \sqrt{2 mE - m^{2} (\omega^{2} - \gamma^{2}) u^{2}} \, {\rm d} u.
\]
It may be noticed that this function is not separable nor $R$-separable.

As a final example we consider the standard Hamiltonian for a simple harmonic oscillator,
\[
H = \frac{p^{2}}{2m} + \frac{m \omega^{2} q^{2}}{2}.
\]
The coordinate transformation
\[
q = \frac{1}{\omega} \sqrt{\frac{2Q}{m}} \, \cos (\omega P), \qquad p = \sqrt{2mQ} \, \sin (\omega P),
\]
is canonical and Eq.\ (\ref{ct}) shows that we can take
\[
K = Q,
\]
with $F = PQ - (Q/\omega) \sin (\omega P) \cos (\omega P)$. Hence, the HJ equation for $K$ is given by
\begin{equation}
Q + \frac{\partial S'}{\partial t} = 0, \label{al}
\end{equation}
whose {\em general}\/ solution is given by $S' = - Qt + f(Q)$, where $f(Q)$ is an {\em arbitrary}\/ function of $Q$ only. Choosing
\[
S' = - Q (t - t_{0}),
\]
where $t_{0}$ is a constant, we obtain a complete solution of the HJ equation (\ref{al}), and from Eq.\ (\ref{sprime}), taking into account that $P = \partial S'/\partial Q = t_{0} - t$, we obtain
\[
S = - \frac{m \omega}{2} q^{2} \tan [\omega (t - t_{0})].
\]
Note that this function is the {\em product}\/ of separated functions of $q$ and $t$.

\section{Concluding remarks}
As shown in Refs.\  2 and 3, and in the example given in Section 2, making use of a constant of motion, one can add a parameter to a given solution of the HJ equation.

The association of the solutions of the HJ equation with certain submanifolds of the extended phase space allows us to readily establish the general relation (\ref{sprime}), avoiding the lengthy computations employed in Refs.\ 2 and 3. This association offers a way of studying the representation of the group of canonical transformations on the principal function $S$ and to understand the structure of the set of solutions of the HJ equation for a given Hamiltonian.

Apart from its intrinsic interest in the Hamiltonian formulation, the results derived here and in Refs.\ 2 and 3 seem relevant in connection with the representation of the canonical transformations in quantum mechanics, owing to the relationship between the HJ equation and the Schr\"odinger equation.

\end{document}